\ifdefined\pdfoutput
\pdfoutput=1
\fi
\documentclass[conference]{IEEEtran}
\IEEEoverridecommandlockouts
\usepackage{cite}
\usepackage{amsmath,amssymb,amsfonts}
\usepackage{algorithmic}
\usepackage{graphicx}
\usepackage{textcomp}
\usepackage{xcolor}
\usepackage{makecell}
\makeatletter
\renewcommand{\maketag@@@}[1]{\hbox{\m@th\normalsize\normalfont#1}}%
\makeatother
\usepackage[caption=false,font=footnotesize]{subfig}
\def\BibTeX{{\rm B\kern-.05em{\sc i\kern-.025em b}\kern-.08em
    T\kern-.1667em\lower.7ex\hbox{E}\kern-.125emX}}
\begin{document}

\title{~\\\LARGE \bf
Real-time Assessment of Distribution Grid Security through Adaptive Smart Meter Measurements

\thanks{This research is supported by the National Science Foundation under Grant. No. ECCS-2045860.}
}
\author{Jiaqi Chen, and Line A. Roald
\thanks{J.~Chen and L.~Roald are with the Department of Electrical and Computer Engineering, University of Wisconsin--Madison, Madison, WI 53706, USA. {\tt\small jiaqi.chen@wisc.edu, roald@wisc.edu}.}%
}


\maketitle
\begin{abstract}
The rapid expansion of distributed energy resources is heightening uncertainty and variability in distribution system operations, potentially leading to power quality challenges such as voltage magnitude violations and excessive voltage unbalance. Ensuring the dependable and secure operation of distribution grids requires system real-time assessment. 
However, constraints in sensing, measurement, and communication capabilities within distribution grids result in limited awareness of the system's state.
To achieve better real-time estimates of distribution system security, we propose a real-time security assessment based on data from smart meters, which are already prevalent in most distribution grids. Assuming that it is possible to obtain a limited number of voltage magnitude measurements in real time, we design an iterative algorithm to adaptively identify a subset of smart meters whose real-time measurements allow us to certify that all voltage magnitudes remain within bounds.
This algorithm iterates between (i) solving optimization problems to determine the worst possible voltage magnitudes, given a limited set of voltage magnitude measurements, and (ii) leveraging the solutions and sensitivity information from these problems to update the measurement set. 
Numerical tests on the IEEE 123 distribution feeder demonstrate that the proposed algorithm consistently identifies and tracks the nodes with the highest and lowest voltage magnitude, even as the load changes over time.

\end{abstract}


\section{Introduction}
The stochastic fluctuations in power injections from distributed energy resources (DERs), alongside load variability, can induce constraint violations within distribution systems \cite{ghijselen2005exact}. These violations may culminate in equipment malfunction, failure of electrical components, and, in extreme circumstances, power outages. Ensuring the dependable and secure operation of distribution grids requires system real-time situational awareness, as this enables distribution system operators (DSOs) to implement appropriate measures to uphold the network's security, reliability, and efficiency without violating operating constraints \cite{zeraati2022meter}. Consequently, there is a pressing need for the development and implementation of more efficient monitoring tools and strategies to ensure the secure operation of distribution networks.
Unfortunately, distribution grids 
typically have very limited
sensing, measurement, and communication capabilities, resulting in limited awareness for the system operator concerning the system state \cite{rasim2018bayesian}. 
Consequently, methods that enable real-time security assessment of the power distribution grid with limited information about the system state, have become a focal point of interest for distribution grid operators.



\subsection{Literature Review}\label{sec:second_subpart}



A fundamental requirement for monitoring distribution grids is the sufficiency and uninterrupted availability of measurement data. Considering the practical scarcity of such data in distribution networks, state estimation (SE) arises as a cost-effective alternative to overcome the lack of measurement data\cite{de2003simulating}. The efficacy of these strategies significantly relies on the volume and accuracy of information, particularly operational data \cite{celli2014dms}. 
The introduction of digital relays, microphasor measurement units ($\mu$PMUs), Intelligent Electronic Devices (IEDs), automated feeder switches, voltage regulators, and smart inverters has provided an opportunity to enhance system observability [2]. However, in practice, only a few real-time measurement devices, such as PMUs, have been installed to monitor medium and low-voltage feeders \cite{zeraati2022meter}, resulting in limited real-time measurements, primarily of current and voltage magnitudes.
To address the challenge of limited real-time sensors, prior approaches to distribution system state estimation (DSSE) have incorporated load forecasts as pseudo-measurements \cite{zamani2017meter,zamzam2020physics}. However, it has been demonstrated in \cite{rasim2018bayesian} that the utilization of forecasted or pseudo-measurements in real time can compromise estimation performance.
Furthermore, due to the increased penetration of renewable energy resources, highly variable loads in the power grid, and the integration of electric vehicles (EVs) into the grid, conventional DSSE methods encounter challenges in effectively managing heightened uncertainty and monitoring dynamic changes and topology configurations\cite{liao2015distribution,menke2019distribution}. Additionally, conventional DSSE approaches, relying on parameters and models of distribution networks, are complex, time-consuming, and highly sensitive to initial conditions \cite{zamzam2019data}. Addressing these challenges has prompted the emergence of intelligent, data-driven, and model-free state estimation algorithms, leveraging artificial intelligence (AI) and machine learning (ML) techniques \cite{rasim2018bayesian,zargar2020multiarea,akrami2021sparse}. These advancements aim to reduce computation time and enhance result accuracy \cite{radhoush2022review}.


The primary objective of DSSE methods is to attain \textit{full observability}  of the system state. 
However, managing significant volumes of data poses a considerable challenge. Moreover, conventional wisdom dictates the necessity of deploying PMUs across the distribution network to obtain accurate state estimates. However,  the implementation of these devices is associated with considerable costs and may require extensive installations to deliver meaningful returns on investment\cite{zamani2017meter}. In contrast, smart meters possess the capability to sense customers' phases, voltage, current, active and reactive powers, and power factors at frequent intervals. Moreover, smart meters are already widely deployed across most distribution networks, thereby eliminating the need for additional overhead in terms of installation and maintenance. Prior work has suggested that such data could be used as near-real-time measurements for inferring system state \cite{alimardani2015distribution}. However, 
at present, limited communication bandwidth poses a challenge in acquiring real-time measurements from all smart meters. Consequently, only a selected subset of smart meter readings can be transmitted to the utility in real time \cite{zeraati2022meter}. 


In order to alleviate the computational burden in DSSE and account for bandwidth considerations associated with smart meter data, our work, in contrast to DSSE, refrains from ensuring \textit{full observability} of the distribution system. Instead, our focus is on enhancing distribution system real-time assessment by identifying the minimum and maximum voltage magnitudes in the feeder, which allows us to achieve the more limited goal of certifying \textit{secure operations}.
\textit{Secure operation} refers to the situation where we can certify that there are no violations of the voltage constraints, despite significant uncertainty regarding the exact power consumption of most loads \cite{molzahn2019grid}.
Specifically, we consider the scenario with specified ranges of load variability (i.e. unknown, but bounded real and reactive power consumption) and a limited subset of nodes where voltages are measured and reported in real time by smart meters, and seek to verify whether there are any voltage violations in the feeder. 

Since the choice of the nodes with real-time measurements impact what we know about the system,  
a specific set of measurement nodes may be unable to certify \textit{secure operations}, while an alternative subset of measurements could potentially ascertain the absence of voltage violations. Consequently, a key contribution of our work is to devise an appropriate strategy for selecting the measurement subset. 

We note that \cite{molzahn2019grid} investigates the certification of secure operation using an optimization-based method given limited measurement and control capabilities, but
does not propose a systematic method for locating the measurement data. In \cite{buason4653282data}, the sensor placement problem is addressed by formulating a bilevel optimization problem, aiming to identify optimal locations for installing sensors capable of capturing all potential violations of voltage magnitude limits across different operating conditions. 
In contrast with \cite{buason4653282data}, which chooses sensor locations and then keeps these locations constant, our work seeks to adaptively update the measurement locations (i.e., which smart meter we obtain data from) in real time based on the current system conditions.

\subsection{Contributions}\label{sec:third_subpart}

In our work, we propose an approach to obtain real-time assessment of distribution grid security through adaptive smart meter measurements. 
We assume that we can access a subset of voltage magnitude measurements in real time (with the number of measurements limited by bandwidth constraints in the smart meter communications infrastructure), that we can adaptively choose which measurements to obtain, and that the power consumption at each node is unknown, but bounded. Given these assumptions, we propose an algorithm that iterates between (i) solving optimization problems to certify whether our current measurement set includes the nodes with the highest and lowest voltage magnitudes in the feeder, and (ii) adapting the measurement set given information obtained from the optimization problem. In each iteration, the algorithm provides bounds on the voltage magnitude ranges in the feeder which can be used to assess distribution grid security. 


The contributions are summarized as follows:

1) We formulate optimization problems to bound the voltage magnitudes in the network by adapting the maximization/minimization framework proposed in \cite{molzahn2019grid} to a setting with a limited set of voltage magnitude measurements. The goal of this step is to determine the extreme achievable voltage magnitudes at each node within the distribution system, given the current set of measurements. 



2) Using the solutions and sensitivities obtained from the optimization problems, we propose a strategy to adjust the measurement set to improve (i.e. narrow) our estimates of the extreme achievable voltage magnitudes.  


3) The effectiveness of the proposed algorithm is validated through numerical demonstrations in the provided case study. The results demonstrate that the algorithm can effectively identify the nodes in the system where the highest/lowest voltages occur and certify that there is indeed no other nodes with worse voltage magnitudes.

The remainder of this paper is organized as follows. Section II describes the formulation of the secure operation certification problem and provides the  approach to update the measurement set to enhance the real-time assessment. In Section III, numerical simulations are conducted to analyze the effectiveness of the proposed method, and conclusions are drawn in Section IV.

\section{Problem Description}\label{sec:Problem Description}

In this section, we first 
adapt the optimization model proposed in \cite{molzahn2019grid}  to a setting with limited voltage measurements. 
Then we propose a strategy to adjust the measurement set to enhance the real-time assessment.

\subsection{Optimization Problem for Extreme Achievable Voltages}\label{sec:first_subpart}

In distribution grids, the power consumption varies due to variability in generation from distributed energy resources (DERs) and load patterns. 
Here, we assume that the exact generation or load is unknown, but bounded to lie within a range of net active and reactive power injections $\boldsymbol {p} \in \begin{bmatrix}\boldsymbol {\underline{p}},\ \boldsymbol {\overline{p}}\end{bmatrix}$,  $ \boldsymbol {q} \in \begin{bmatrix}\boldsymbol {\underline{q}},\ \boldsymbol {\overline{q}}\end{bmatrix}$. Here $\boldsymbol {p}, \boldsymbol {q}$ represent vectors of real and reactive power injections, while $\boldsymbol {\underline{p}}$, $\boldsymbol {\overline{p}}$, $\boldsymbol {\underline{q}}$, and $\boldsymbol {\overline{q}}$ are vectors of their respective upper and lower bounds. 

Due to the bandwidth limitations in the smart meter communications infrastructure, the number of nodes where we have access to the real-time voltage magnitude is limited. We use the set $\mathcal{V}$ to represent the subset of nodes from which we obtain voltage magnitude data in real time. The number of such nodes $|\mathcal{V}|$ is limited due to the bandwidth, but we can choose which nodes are contained in $\mathcal{V}$.

Our objective is to determine whether there exists a load realization $\boldsymbol {p}, \boldsymbol {q}$ in the specified ranges that causes voltage violations at any nodes, given the measured voltage magnitudes at nodes in $\mathcal{V}$. 
In contrast to DSSE, which estimates the voltage magnitude for each node, our (more limited) objective is to determine the \textit{extreme achievable voltage magnitudes}. 
Once we have bounds on the extreme achievable voltages, we can compare these values with the upper and lower engineering limits to ascertain whether there is a potential violation of voltage constraints\footnote{We note that while our focus is primarily on voltage magnitude constraints, which are typically of greatest concern in distribution grid operations \cite{8520738}, a similar methodology could be applied to investigate other factors such as current flow or voltage unbalance.}.


The secure operation
certification problem is formulated as the following optimization problem, aiming to find the extreme achievable voltage magnitude at all nodes subject to constraints representing the operation of a three-phase distribution system:
\begin{subequations}
    \label{eq:smartmodel}
    \begin{align}
        \text{(P1)}\quad\quad &\underline{u}^{\phi}_{n}=\operatorname*{min}_{\boldsymbol p, \boldsymbol q, \boldsymbol u}{u}^{\phi}_{n}\quad\mathbf{or}\quad\overline{u}^{\phi}_{n}=\operatorname*{max}_{\boldsymbol p, \boldsymbol q, \boldsymbol u}{u}^{\phi}_{n}\label{eq:smartmodel1}\\
        \text{s.t.}\quad\quad & (\forall n\in\mathcal{N})\notag\\
        &\underline{p}^{\phi}_{k}\leq p^{\phi}_{k}\leq\overline{p}^{\phi}_{k}, \quad\quad \forall k\in\mathcal{N}\backslash ref,\quad\label{eq:smartmodel2}  \\
&\underline{q}^{\phi}_{k}\leq q^{\phi}_{k}\leq\overline{q}^{\phi}_{k},\quad\quad\forall k\in\mathcal{N}\backslash ref,\label{eq:smartmodel3} \\
&{u}^{\phi}_i={u}^{{\text {measure}},\phi}_i,\quad\quad \forall i\in\mathcal{V}, \quad(\lambda^{\phi}_i),\label{eq:smartmodel4}\\
&{u}^{\phi}_{\text {root }}={u}^{\phi}_{\text {ref}},\label{eq:smartmodel5}\\
&p_{ij}^{{\phi}}=\sum_{k\in N_{d}\left(j\right)}p_{jk}^{\phi}+p_{j}^{\phi},\label{eq:smartmodel6}\\
&q_{ij}^{{\phi}}=\sum_{k\in N_{d}\left(j\right)}q_{jk}^{\phi}+q_{j}^{\phi},\label{eq:smartmodel7}\\
&u_{i}^{{\phi}}-u_{j}^{{\phi}}=2\cdot\left(r_{ij}^\phi\cdot p_{ij}^\phi+x_{ij}^\phi\cdot q_{ij}^\phi\right). \label{eq:smartmodel8}
\end{align}
\end{subequations}

The objective function \eqref{eq:smartmodel1} minimizes or maximizes the square of voltage magnitude achievable at a particular node $n\in\mathcal{N}$, where $\underline{u}^{\phi}_{n}$ and $\overline{u}^{\phi}_{n}$ are the minimum and maximum achievable squared voltage magnitude at phase $\phi$ at node $n$, ${u}^{\phi}_{n}$ is the square of voltage magnitude at phase $\phi$ at node $n$, $\mathcal{N}$ is the set of all the nodes in the distribution system, and $\boldsymbol{p}, \boldsymbol{q}$ and $\boldsymbol{u}$ are the vectors of three-phase active power injection, reactive power injection and square of voltage magnitude.

Constraints \eqref{eq:smartmodel2} and \eqref{eq:smartmodel3} model the range of variability in the net power injections, where $p^{\phi}_{k}$ and $q^{\phi}_{k}$ are the active power injection, and reactive power injection at phase $\phi$ at all nodes except for the reference node. 

Constraint \eqref{eq:smartmodel4} enforces that the square of voltage magnitudes equals the measured value if the smart meter at the respective node is selected to report its data in real time. Here, $\mathcal{V}$ denotes the subset of nodes where smart meters are chosen to provide real-time measured voltage magnitude to the operator.

Constraints \eqref{eq:smartmodel5} to \eqref{eq:smartmodel8} represent three-phase power flow constraints. The power flow with a radial structure can be described using the branch power flow model. We adopt a linear distribution power flow (LinDistFlow) model \cite{baran1989optimal2} to construct a linear optimization model. Here, ${u}^{\phi}_{\text {root}}$ and ${u}^{\phi}_{\text {ref}}$ represent the squared root node voltage magnitudes and the squared reference voltage magnitudes at phase $\phi$, while $p_{ij}^{{\phi}}$ and $q_{ij}^{{\phi}}$ are the branch active power flow and reactive power flow on line $ij$ at phase $\phi$, respectively. $N_{d}\left(j\right)$ is the set of nodes connected by the lateral branching out from node $j$.


\subsection{Criteria for Success}
\label{sec:criteria_success}
Based on the problem above, we define two criteria for success of our problem.

\subsubsection{Successful Certification for Secure Operations}
Our first criterion for success is whether we can verify the absence of voltage violations given the range of load variability \eqref{eq:smartmodel2} and \eqref{eq:smartmodel3} and the set of measurement nodes $\mathcal{V}$. We use $\boldsymbol {u}^{\text{min}}, \boldsymbol {u}^{\text{max}}$ to represent the lower and upper voltage engineering limits corresponding to secure operating ranges that adhere to power quality and safety requirements. We propose the following criterion to certify the \textit{secure operation}, 
\begin{equation}
\label{eq:smart_secu}
\begin{aligned}
\text{\textbf{Criterion 1}:} \;\;{u}^{\phi, \text{min}}_{i} \leq \underline{u}^{\phi}_{i} \;\;\text {and} \;\;\overline{u}^{\phi}_{i} \leq {u}^{\phi, \text{max}}_{i}, \;\forall i \in \mathcal{N}.
\end{aligned}
\end{equation}

If the minimum and maximum achievable voltage magnitudes ($\underline{u}^{\phi}_{i}$ and $\overline{u}^{\phi}_{i}$) obtained from 
(P1)
fall within the specified engineering limits $\boldsymbol {u}^{\text{min}}, \boldsymbol {u}^{\text{max}}$ for all nodes, 
the power flow solutions associated with all potential power injections within the specified ranges \eqref{eq:smartmodel2} and \eqref{eq:smartmodel3} will adhere to the engineering limits on voltage magnitudes, provided we have real-time measured voltage magnitude data of the subset nodes $\mathcal{V}$. In this case, we can guarantee that the system is operating securely.

\subsubsection{Successful Identification of Highest and Lowest Voltages}
Our second criterion for success is whether we can verify that the measurement set $\mathcal{V}$ contains the nodes with the highest and lowest voltage magnitudes in the feeder. 
We consider the problems (P1) with the most extreme achievable voltage magnitudes, i.e. the node with the lowest achievable voltage $\underline{u}^{\phi}_{i}$, which we will refer to as $\underline{u}^{\phi,\text{lowest}}$, and the node with the largest achievable voltage magnitude $\overline{u}^{\phi}_{i}$, which we will refer to as $\overline {u}^{\phi, \text{largest}}$. We also identify the highest and lowest measured voltage magnitudes ${\underline{u}}^{\text{measure}},\overline{u}^{{\text {measure}}}$ among our measured nodes $\mathcal{V}$. We then calculate the differences $\underline{\Delta u},\overline{\Delta u}$ between the highest (and lowest) achievable voltage magnitudes, as identified by the optimization problems, and the highest (and lowest) measured voltage magnitudes,
\begin{equation}
\label{eq:smart_difference_lower}
\begin{aligned}
\underline{\Delta u}=\underline{u}^{{\text {measure}}} -\underline{u}^{\phi,\text{lowest}} ,
\end{aligned}
\end{equation}
\begin{equation}
\label{eq:smart_difference_upper}
\begin{aligned}
\overline{\Delta u}=\overline{u}^{\phi, \text{largest}}-\overline{u}^{{\text {measure}}}.
\end{aligned}
\end{equation}
If the difference $\underline{\Delta u},\overline{\Delta u}$ is equal or less than a predefined small enough value $\varepsilon$, we can certify that we have found the nodes in the network with the highest and lowest voltage magnitudes (within the tolerance $\varepsilon$). We also know that it is not possible to identify a measurement set $\mathcal{V}$ that would lead to tighter bounds on the minimum and maximum achievable voltage magnitudes $\underline{u}^{\phi,\text{lowest}}$ $\overline{u}^{\phi, \text{largest}}$, as those bounds are already equal to actual measured voltages.

\subsection{Strategy for Adapting the Measurement Set}\label{sec:3rd_subpart}

The extreme achievable voltages $\underline{u}^{\phi}_{i}$ and $\overline{u}^{\phi}_{i}$ depend on the considered range of load variability \eqref{eq:smartmodel2} and \eqref{eq:smartmodel3} and the subset of nodes $\mathcal{V}$ where we can access real-time voltage magnitudes. 
While we consider the range of load variability to be an input to our problem, we can choose the set of nodes $\mathcal{V}$ which provides the \textit{narrowest range of extreme achievable voltages}.
Specifically,
we aim to choose $\mathcal{V}$ in a way that maximizes the lowest achievable voltages $\underline{u}^{\phi,\text{lowest}}_{i}$ and minimizes the largest extreme achievable voltage $\overline{u}^{\phi,\text{largest}}_{i}$ to let $\underline{u}^{\phi,\text{lowest}}_{i}$, $\overline{u}^{\phi,\text{largest}}_{i}$ become as close as possible to the actual highest and lowest voltage in the system, and ideally make them equal (as outlined in our second criterion for success above).

The challenge in updating the measurement set arises from the fact that we do not know the new voltage measurement until we have measured it. 
Assuming that our sampling procedure allows us to update the measurement set $\mathcal{V}$ and obtain a new measurement before the load changes significantly (e.g. within a few seconds or a minute), 
we propose to update $\mathcal{V}$ for the next time step based on the solutions and sensitivity information obtained from our previous solution of 
(P1). 

Our goal is to change our measurement set by ``move in'' (i.e. adding a measurement to) the node that has the most impact on the values of the lowest and largest extreme achievable voltages ($\underline{u}^{\phi,\text{lowest}}$, $\overline {u}^{\phi, \text{largest}}$) of the problem (P1) and ``remove'' (i.e. stop taking a measurement from) the node with the least impact.
Since the dual variable of a constraint represents the impact of this constraint on the objective function, we use the dual variable $\lambda^{\phi}_i$ of the constraint \eqref{eq:smartmodel4} to analyze the impact of the measured nodes on the values of extreme achievable voltages ($\underline{u}^{\phi}_{i}$ and $\overline{u}^{\phi}_{i}$).


\subsubsection{Metrics to assess the value of a measured node} We define two metrics to help us assess the value (i.e. the information provided by) different nodes in our measurement set. For each measured node, we define:  

\noindent\textit{\underline{Metric I:}}
We consider the two problems (P1) which have the lowest $\underline{u}^{\phi}_{i}$ and largest $\overline{u}^{\phi}_{i}$ (i.e. the problems that provide $\underline{u}^{\phi,\text{lowest}}$ and $\overline {u}^{\phi, \text{largest}}$). We take the absolute value of the dual variable of the voltage measurement constraint (1d) $\lambda^{\phi}_i$ for each measured node $i\in\mathcal{V}$ as our Metric I, 
\begin{equation}
\underline{\lambda}^{\text{\uppercase\expandafter{\romannumeral1}}}\!= \!\left\{|\lambda_{i,k}^{\phi}| \mid \forall i\in\mathcal{V},\phi \!\in \!\{a, b, c\}\right\},
\end{equation}
\begin{equation}
\overline{\lambda}^{\text{\uppercase\expandafter{\romannumeral1}}}\!= \! \left\{|\lambda_{i,\ell}^{\phi}| \mid \forall i\in\mathcal{V}, \phi \!\in \!\{a, b, c\}\right\}.
\end{equation}

Here, subscript $k$ refers to the problem (P1) for the node with the lowest achievable voltage and $\ell$ refers to the problem (P1) for the node with the highest achievable voltage.
\noindent\textit{\underline{Metric II:}}
We consider the average impact of the measured nodes on the extreme achievable voltage magnitude across all problems (P1). However, since most nodes only provide information regarding other, nearby nodes (e.g. nodes on the same branch of the feeder), we introduce a special way of computing the average that only considers nodes for which the impact (as measured by the value of the dual variable associated with constraint (1d)) exceeds a given threshold $\lambda^{\text{threshold}}$. This ensures that we not only choose nodes that are located in the longest feeder branch. 
We define the set of nodes $\mathcal{N}_i$ as those for which the absolute value of the dual variable associated with the measurement at node $i$ exceeds the threshold ($|\lambda_{i,k}|>\lambda^{\text{threshold}}$), where $k$ represents the problem (P1) for node $k$. 
We then define Metric II as
\vspace{-4pt}
\begin{equation}
{\lambda}^{\text{\uppercase\expandafter{\romannumeral2}}}\!= \!\left\{\text{mean} |\lambda_{i,k}^{\phi}| \mid \forall k\in\mathcal{N}_i, \phi \!\in\! \{a, b, c\}\right\}.
\vspace{-3pt}
\end{equation}

\subsubsection{Avoiding recursive loops}
One important challenge of our approach is to prevent recursive loops, where nodes are switched in a cyclic manner (e.g., switching from node A to node B in iteration 1 and then reverting from node B to node A in iteration 2, and so forth). To address this, we implement a \textit{loop check} after we choose a new measurement set $\mathcal{\hat V}$. We define \textit{the existence of a loop} if the new measurement set is the same as one of the past \textit{M} measurement sets $\mathcal{V}$, where \textit{M} is a predefined value. If a loop is identified, we select another new measurement set $\mathcal{\hat V}$ to avoid the loop, which will be illustrated in detail in the following strategy. 

\subsubsection{Iterative Algorithm to Adjust the Measurement Set}
At each time step, we use the following algorithm to adjust the measurement set. Assuming that we exchange one node at a time, the goal is to find a new subset $\mathcal{\hat V}$ to achieve a larger $\underline{u}^{\phi,\text{lowest}}$ or a lower $\overline {u}^{\phi, \text{largest}}$ to narrow the the extreme achievable range for a better real-time assessment.

\noindent\textit{\underline{Step 1:}}
In the first step, we solve the secure operation certification problem (P1) for all the nodes and phases. 
Check if \textbf{Criterion 1} is satisfied; if yes, we can guarantee that the system is operating securely.

\noindent\textit{\underline{Step 2:}}
In the second step, we check if we need to update the measurement set by checking if $\underline{\Delta u}\leq\varepsilon,\overline{\Delta u}\leq\varepsilon$. 
If these conditions are satisfied, we have already identified the nodes with the highest lowest voltage, and it is not possible to further improve the set of measurements to achieve a tighter range. In this case, we keep the measurement set the same and move back to Step 1 once we receive new measurements.  Moreover, if \textbf{Criterion 1} is not satisfied at this time, we can confirm that the system is operating insecurely.

If the differences $\underline{\Delta u},\overline{\Delta u}$ are larger than $\varepsilon$, it may be possible to achieve a better bound on $\underline{u}^{\phi,\text{lowest}}$ and $\overline{u}^{\phi, \text{largest}}$, by updating the measurement set. In this case, we continue to Step 3.

\noindent\textit{\underline{Step 3 (Identify ``Move in'' node):}}
Because we want to know if the node with the lowest or largest extreme achievable voltage value $\underline{u}^{\phi,\text{lowest}}, \overline{u}^{\phi, \text{largest}}$ really has such a low or large voltage value, we select this node to be added to the new measurement set $\mathcal{\hat V}$ to obtain an exact real-time measurement on this node. 

\vspace{-1.3mm}
\noindent\textit{\underline{Step 4 (Identify ``Remove'' node):}}
To identify the node we want to remove, we use Metric I $(\underline{\lambda}^{\text{\uppercase\expandafter{\romannumeral1}}}$, $\overline{\lambda}^{\text{\uppercase\expandafter{\romannumeral1}}})$ and Metric II $({\lambda}^{\text{\uppercase\expandafter{\romannumeral2}}})$ to choose the node that should be removed. 
Assuming that the number of measurements 
per phase is fixed, we remove a node from the same phase where we want to add a node (i.e. the phase with the lowest or largest extreme achievable voltage value $\underline{u}^{\phi,\text{lowest}}, \overline{u}^{\phi, \text{largest}}$). Therefore, we compare Metric I and II for the nodes in that phase. We follow a hierarchical approach to decide which node to remove.

(i) First, we consider the measured nodes that have the least impact on $\underline{u}^{\phi,\text{lowest}}$ or $\overline {u}^{\phi, \text{largest}}_{i}$ by looking at Metric I. We compare $\underline{\lambda}^{\text{\uppercase\expandafter{\romannumeral1}}}$ and $\overline{\lambda}^{\text{\uppercase\expandafter{\romannumeral1}}}$ of each measured node and remove the node with the smallest value from $\mathcal{V}$. 

(ii) If multiple nodes have the same smallest value for Metric I, we remove the node that has the least average impact. Specifically, we choose the node with the smallest value for Metric II ${\lambda}^{\text{\uppercase\expandafter{\romannumeral2}}}$ among the nodes that have the same smallest Metric I, and remove this node from $\mathcal{V}$. 

\noindent\textit{\underline{Step 5 (Loop check):}}
Next, we do a \textit{loop check}. If \textit{the loop exists}, according to the criterion defined above, we exclude the previously identified node as a possible ``Remove'' node and go back to Step 4. In this case, we turn to remove
the node with the next smallest Metric value. 

If no \textit{loop exists}, we have identified a new measurement set $\mathcal{\hat V}$ and return to Step 1 once we obtain the new measurements. 

\begin{figure}
    \centering
    \includegraphics[width=0.33\textwidth]{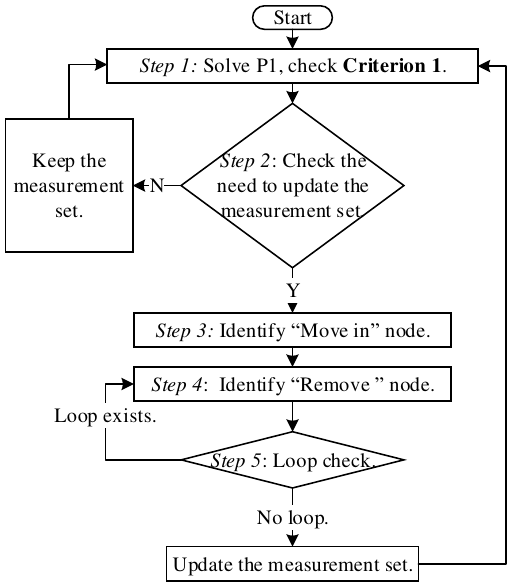}
    \caption{Flow chart of the proposed strategy at one time step.}
    \label{fig:flowchart}
    \vspace{-15pt}
\end{figure}

The steps of the algorithm are summarized in the flowchart in Fig. \ref{fig:flowchart}. Throughout the iterative process, the proposed method enhances grid reliability by enabling the utility to either certify the secure operation of the system or detect potential operational issues. Since the secure operation certification problem (P1) is linear and Step 1 can be accelerated through parallel computation, the computational complexity of the proposed algorithm is low, making it applicable to large systems. 

\vspace{-10pt}

\section{Numerical Test}

\subsection{Simulation Setup}
The proposed approach is tested on a modified unbalanced three-phase 123-bus distribution system \cite{zheng2018distributed}. We simulate the load change by multiplying the given load consumption with a factor generated based on the real load data of the oak-park substation located in a suburb of Portland, OR. The information of branch parameters and load profiles is available online\cite{CAISO_FRP:1}. Considering the bandwidth limitations
in the smart meter communications infrastructure, we assume for each phase, the number of measurements \textit{K} is fixed and we can switch one measurement on one phase at one time step. The power flow calculated with the LinDistFlow method is used as the measured value for the measured nodes. We check the loop in the past 20 time steps (\textit{M} = $20$). The predefined value $\varepsilon$ is set to be $1e^{-4}$ and the threshold $\lambda^{\text{threshold}}$ is set to be $0.3$. These values are found to be a reasonable choice based on experience from running numerical simulations. We test the approach with different initial measurement sets, different fixed measurement numbers \textit{K}, and different load profiles.  
All simulations were implemented using MATLAB on a personal laptop with an Apple M1 Pro processor and $32 \mathrm{~GB}$ of RAM.

\subsection{Test Results}

\begin{figure}
    \centering
    \includegraphics[width=0.4\textwidth]{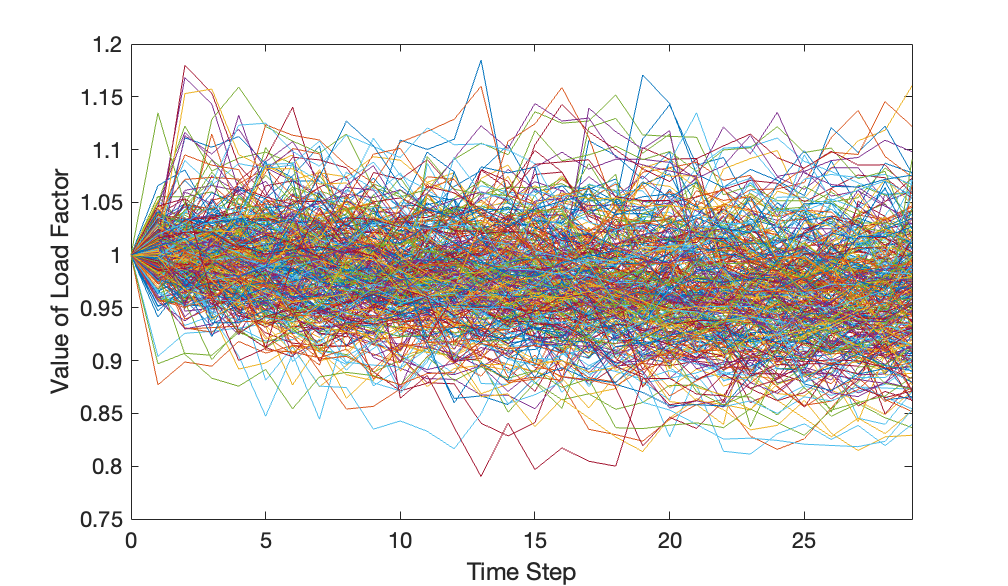}
    \caption{Value of the factor that multiplies with the given load on scenario 1.}
    \label{fig:load_scenario1}
    \vspace{-10pt}
\end{figure}

First, we test the approach in scenario 1 with the initial measurement set on nodes $\{22,23,24,26,27,28,29,30,31,32\}$ at each phase with the fixed measurement number \textit{K} = $10$. The real-time assessment process started at $00:00$ AM of one day and the time step to update the measurement set is 1 minute. Because in scenario 1 there are no distributed generations, we focus on the lower bound of the voltage magnitude. Due to the space limitation, we only illustrate the result of the first 30 time steps in Table \ref{tab:RESULT OF SCENARIO 1}. The corresponding factor that multiplies with the given load is illustrated in Fig. \ref{fig:load_scenario1}. 


\begin{table*}[htbp]
  \centering
  \caption{Result of Scenario 1}
    \begin{tabular}{cc|ccc|ccc|ccc|cc}
\hline
\hline
          \multicolumn{1}{c}{}&\multicolumn{1}{c}{}       & \multicolumn{3}{c}{Lowest Measured Voltage} & \multicolumn{3}{c}{Lowest Achievable Voltage} & \multicolumn{3}{c}{Lowest Voltage} &      \multicolumn{2}{c}{Voltage Difference} \\
          \cline { 3 - 13 }
Time & Status & Node  & $\phi$ & $u$  & Node  & $\phi$ & $u$ & Node  & $\phi$ & $u$  & $\underline{\Delta u}$ & $\underline{\Delta u}^{\text{actual}}$ \\
\hline
    0& 0& 30,31 &$a$ & 0.9140  & 115   &$a$  & 0.7413  & 115   & $a$ & 0.8605  & 1.73e-1 & 1.19e-1 \\

    1& 0& 115   & $a$ & 0.8601  & 86    &$c$& 0.8044  & 115   & $a$ & 0.8601  & 5.57e-2 & 5.57e-2 \\

    2& 0& 115   & $a$ & 0.8604  & 97    & $b$& 0.8135  & 115   & $a$ & 0.8604  & 4.69e-2 & 4.69e-2 \\

    3& 0& 115   & $a$ & 0.8610  & 95    & $a$ & 0.8290  & 115   &$a$ & 0.8610  & 3.20e-2 & 3.20e-2 \\

    4& 0& 115   &$a$ & 0.8604  & 67    &$c$& 0.8420  & 115   &$a$ & 0.8604  & 1.84e-2 & 1.84e-2 \\

    5& 0& 115   &$a$ & 0.8601  & 72    &$a$ & 0.8480  & 115   &$a$ & 0.8601  & 1.21e-2 & 1.21e-2 \\

    6& 0& 115   &$a$ & 0.8612  & 83    &$a$ & 0.8578  & 115   &$a$ & 0.8612  & 3.41e-3 & 3.41e-3 \\

    7& 0& 115   &$a$ & 0.8621  & 112   &$a$ & 0.8595  & 115   &$a$ & 0.8621  & 2.62e-3 & 2.62e-3 \\

    8& 0& 115   &$a$ & 0.8626  & 72    &$a$ & 0.8524  & 115   &$a$ & 0.8626  & 1.02e-2 & 1.02e-2 \\

    9& 0& 112   &$a$ & 0.8662  & 115   &$a$ & 0.8556  & 115   &$a$ & 0.8634  & 1.06e-2 & 7.85e-3 \\

    10    & 2& 115   &$a$ & 0.8634  & 114   &$a$ & 0.8634  & 115   &$a$ & 0.8634  & 0& 0 \\

    11    & 2& 115   &$a$ & 0.8631  & 114   &$a$ & 0.8631  & 115   &$a$ & 0.8631  & 0& 0 \\

    12    & 0& 115   &$a$ & 0.8639  & 105   &$c$& 0.8626  & 115   &$a$ & 0.8639  & 1.32e-3 & 1.32e-3 \\

    13    &2& 115   &$a$ & 0.8639  & 114   &$a$ & 0.8639  & 115   &$a$ & 0.8639  & 0& 0 \\

    14    &1 & 115   &$a$ & 0.8650  & 115   &$a$ & 0.8650  & 115   &$a$ & 0.8650  & 0& 0 \\

    15    &1 & 115   &$a$ & 0.8650  & 115   &$a$ & 0.8650  & 115   &$a$ & 0.8650  & 0& 0 \\

    16    &1 & 115   &$a$ & 0.8655  & 115   &$a$ & 0.8655  & 115   &$a$ & 0.8655  & 0& 0 \\

    17    &1 & 115   &$a$ & 0.8646  & 115   &$a$ & 0.8646  & 115   &$a$ & 0.8646  & 0& 0 \\

    18    & 2& 115   &$a$ & 0.8646  & 114   &$a$ & 0.8646  & 115   &$a$ & 0.8646  & 0& 0 \\

    19    &1 & 115   &$a$ & 0.8651  & 115   &$a$ & 0.8651  & 115   &$a$ & 0.8651  & 0& 0 \\

    20    &1 & 115   &$a$ & 0.8653  & 115   &$a$ & 0.8653  & 115   &$a$ & 0.8653  & 0& 0 \\

    21    & 2& 115   &$a$ & 0.8659  & 114   &$a$ & 0.8659  & 115   &$a$ & 0.8659  & 0& 0 \\

    22    & 2& 115   &$a$ & 0.8663  & 114   &$a$ & 0.8663  & 115   &$a$ & 0.8663  & 0& 0 \\

    23    & 2& 115   &$a$ & 0.8659  & 114   &$a$ & 0.8659  & 115   &$a$ & 0.8659  & 0& 0 \\

    24    &1 & 115   &$a$ & 0.8659  & 115   &$a$ & 0.8659  & 115   &$a$ & 0.8659  & 0& 0 \\

    25    &1 & 115   &$a$ & 0.8653  & 115   &$a$ & 0.8653  & 115   &$a$ & 0.8653  & 0& 0 \\

    26    & 2& 115   &$a$ & 0.8660  & 114   &$a$ & 0.8660  & 115   &$a$ & 0.8660  & 0& 0 \\

    27    & 2& 115   &$a$ & 0.8661  & 114   &$a$ & 0.8661  & 115   &$a$ & 0.8661  & 0& 0 \\

    28    & 2& 115   &$a$ & 0.8665  & 114   &$a$ & 0.8665  & 115   &$a$ & 0.8665  & 0& 0 \\

    29    &1 & 115   &$a$ & 0.8661  & 115   &$a$ & 0.8661  & 115   &$a$ & 0.8661  & 0& 0 \\

\hline
\hline
    \end{tabular}%
  \label{tab:RESULT OF SCENARIO 1}%
\end{table*}%

In Table \ref{tab:RESULT OF SCENARIO 1}, the first column represents the time step, while the second column represents the status of the algorithm at this time step. Status 0 is used to indicate that the difference $\underline{\Delta u}$ between the lowest achievable voltage value 
$\underline{u}^{\phi,\text{lowest}}$ and the lowest measured voltage value $\underline{u}^{{\text {measure}}}$ is larger than the predefined value $\varepsilon$, i.e., when the status is 0, it is necessary to update the set of measurements. Status 1 is used to indicate a situation in which the node with the lowest achievable voltage value $\underline{u}^{\phi,\text{lowest}}$ is the same as the node with the lowest measured voltage value $\underline{u}^{{\text {measure}}}$, i.e., status 1 indicates that we have found (and measured) the node with the lowest voltage magnitude, so we don't need to update the measurement set. Status 2 indicates a situation in which the difference $\underline{\Delta u}$ is equal or less than the value $\varepsilon$ but the node with the lowest achievable voltage value $\underline{u}^{\phi,\text{lowest}}$ is not the same as the node with the lowest measured voltage value $\underline{u}^{{\text {measure}}}$. Thus, with status 2, we also do not need to update the measurement set. The next columns in Table I show the lowest measured voltage (i.e. the lowest known voltage), the lowest achievable voltage (as indicated by the optimization problems) and the actual lowest voltage (which generally wouldn't be known). The last two columns illustrate the difference of the square of voltage magnitude, where $\underline{\Delta u}$ represents the difference between the lowest achievable voltage value $\underline{u}^{\phi,\text{lowest}}$ and the lowest measured voltage value $\underline{u}^{{\text {measure}}}$,  $\underline{\Delta u}^{\text{actual}}$ represents the difference between the lowest achievable voltage value $\underline{u}^{\phi,\text{lowest}}$ and the system's actual lowest voltage value ${u}^{{\text {actual},\text{lowest}}}$.

We can observe that on status 1 and 2, when the lowest achievable voltage $\underline{u}^{\phi,\text{lowest}}$ is the same as the lowest measurement voltage value $\underline{u}^{{\text {measure}}}$ (within the tolerance $\varepsilon$), i.e., $\underline{\Delta u}\leq\varepsilon$, the lowest achievable voltage $\underline{u}^{\phi,\text{lowest}}$ is also always the same as the system's actual lowest voltage ${u}^{{\text { actual},\text{lowest}}}$ (within the tolerance $\varepsilon$), i.e., $\underline{\Delta u}^{\text{actual}}\leq\varepsilon$. This is because the lowest achievable voltage $\underline{u}^{\phi,\text{lowest}}$ is always less than all the system real operating voltage, and if the lowest achievable voltage $\underline{u}^{\phi,\text{lowest}}$ reaches the lowest measured voltage value, (which is an actual voltage value), i.e., status 1 or 2, it reaches the real lowest voltage value. Therefore, on status 1 and 2, the system's highest and lowest voltage values and the nodes where they occur are the same as those of the highest and lowest measured voltages. In this case, we obtain a \textit{good system real-time assessment}.

It can be observed that with the initial measurement set, it takes 11 time steps to find a \textit{good measurement set} where the lowest achievable voltage $\underline{u}^{\phi,\text{lowest}}$ is almost the same as the actual lowest voltage value ${u}^{{\text {actual},\text{lowest}}}$, i.e. status 2 at time 10, which indicates a \textit{good system real-time assessment}. The value of the lowest achievable voltage $\underline{u}^{\phi,\text{lowest}}$ shows an overall upward trend from time 0 to time 10. After that, since the load does not change significantly, we do not update the measurement set except at time 12. We observe that after we get the \textit{good system real-time assessment} at time 10, a \textit{good system real-time assessment} (status 1 or 2) is consistently achieved in most subsequent time steps. Moreover, we observe that although at time 12 we don't have a \textit{good system real-time assessment}, we can find a new \textit{good measurement set} in just one iteration. This is because the initial measurement set at time 12 is superior to that at time 0. 

To further illustrates the measurement update process, we use Table \ref{tab:Measurement Update Process} to show the update process from time 12 to time 14. Due to space limitations, we only show the information of the phase where the lowest achievable voltage value $\underline{u}^{\phi,\text{lowest}}$ occurs. It can be observed that at time 12, we solve the problem (P1) with the measurement set of phase $c$ on nodes $\{22,86,24,26,27,67,29,30,31,32\}$, the lowest achievable voltage value $\underline{u}^{\phi,\text{lowest}}$ occurs on node $105$ on phase $c$. The voltage difference $\underline{\Delta u}$ is larger than the value $\varepsilon$. Therefore, we move node $105$ on phase $c$ into the new measurement set $\mathcal{\hat V}$. Next, we compare Metric I on phase $c$ and find the smallest value of Metric I is $0$ and occurs on nodes $\{29, 30,31\}$. Therefore, we compare Metric II of nodes $\{29, 30,31\}$ and find node 29 has the smallest Metric II value. So we remove node 29 on phase $c$ from the measurement set and check that there no \textit{loop exists}. The updated new measurement set $\mathcal{\hat V}$ is nodes $\{22,86,24,26,27,67,105,30,31,32\}$ on phase $c$. At time 13 and time 14, we take \textit{Step 1} and \textit{Step 2} and find the voltage difference $\underline{\Delta u}$ is 0, which is less than the value $\varepsilon$. Therefore, we don't update the measurement set. The difference between the scenarios at time 13 and time 14 is that at time 14, the node with lowest achievable voltage value $\underline{u}^{\phi,\text{lowest}}$ (node $115$) is already in the measurement set, i.e., status 1.

\begin{table*}[htbp]
  \centering
  \caption{Measurement Update Process}
    \begin{tabular}{rcccc|cccc|cccc}
    \hline
\hline
          & \multicolumn{4}{c}{Time 12 Phase $c$}   & \multicolumn{4}{c}{Time 13 Phase $a$}   & \multicolumn{4}{c}{Time 14 Phase $a$} \\
          \hline
          & Start Set & Mertic I & Metric II & New Set & Start Set & Mertic I & Metric II & New Set & Start Set & Mertic I & Metric II & New Set \\
          \hline
          & 22    & 0.07740  & 3.02490  & 22    & 22    & 1.48e-16 & 5.18140  & 22    & 22    & 0     & 5.18044  & 22 \\
          & 86    & 0.78066  & 0.81971  & 86    & 72    & 2.66e-17 & 0.49237  & 72    & 72    & 0     & 0.49237  & 72 \\
          & 24    & 0.04222  & 2.98695  & 24    & 24    & 2.35e-16 & 7.13266  & 24    & 24    & 0     & 7.13392  & 24 \\
          & 26    & 0.00304  & 6.19873  & 26    & 95    & 8.33e-17 & 0.96107  & 95    & 95    & 0     & 0.96107  & 95 \\
          & 27    & 0.00468  & 1.04175  & 27    & 112   & 0     & 0.63394  & 112   & 112   & 0     & 0.63394  & 112 \\
          & 67    & 0.16017  & 0.56466  & 67    & 28    & 2.78e-17 & 1.28261  & 28    & 28    & 0     & 1.28280  & 28 \\
          & 29    & 0     & 0     & 105   & 29    & 6.61e-17 & 3.09891  & 29    & 29    & 0     & 3.09936  & 29 \\
          & 30    & 0     & 4.95827  & 30    & 83    & 0     & 0.87210  & 83    & 83    & 0     & 0.87210  & 83 \\
          & 31    & 0     & 3.23078  & 31    & 31    & 1.39e-17 & 1     & 31    & 31    & 0     & 1     & 31 \\
          & 32    & 0.00164242 & 2.33333  & 32    & 115   & 1     & 0.48570  & 115   & 115   & 1     & 0.48570  & 115 \\
          \hline
          \hline
          & \makecell{Node of\\  $\underline{u}^{\phi,\text{lowest}}$} & \makecell{Phase of\\  $\underline{u}^{\phi,\text{lowest}}$}& $\underline{u}^{\phi,\text{lowest}}$ &   & \makecell{Node of\\  $\underline{u}^{\phi,\text{lowest}}$} & \makecell{Phase of\\  $\underline{u}^{\phi,\text{lowest}}$}& $\underline{u}^{\phi,\text{lowest}}$ &   & \makecell{Node of\\  $\underline{u}^{\phi,\text{lowest}}$} & \makecell{Phase of\\  $\underline{u}^{\phi,\text{lowest}}$}  & $\underline{u}^{\phi,\text{lowest}}$ &   \\
          \hline
          & 105   &  $c$    & 0.86256  &      & 114   & $a$     & 0.86394  &      & 115   &   $a$   & 0.86498  &  \\
          \hline
\hline
    \end{tabular}%
  \label{tab:Measurement Update Process}%
\end{table*}%

\begin{figure}
    \centering
    \includegraphics[width=0.4\textwidth]{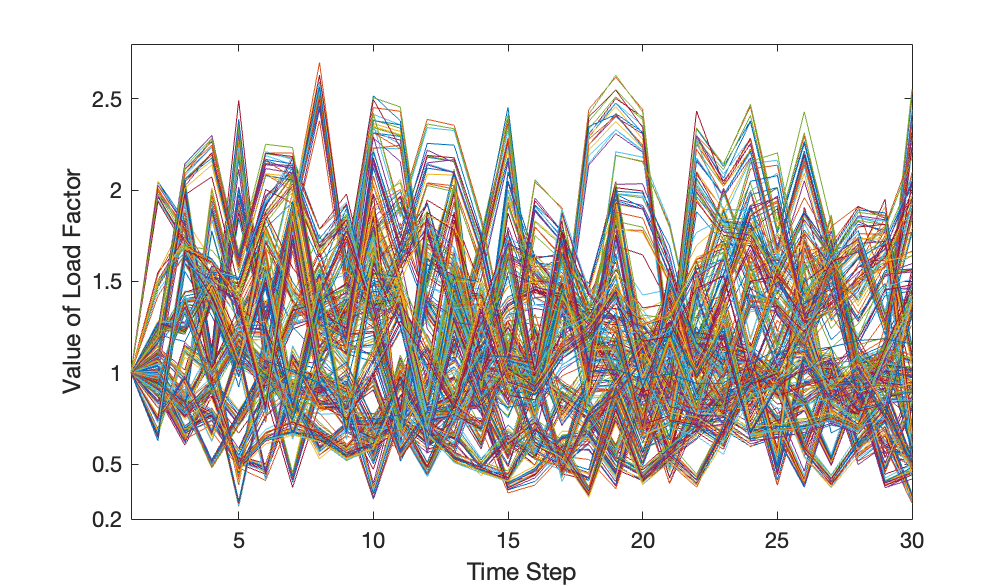}
    \caption{Value of the factor that multiplies with the given load on scenario 5.}
    \label{fig:load_scenario5}
    \vspace{-10pt}
\end{figure}

Next, we analyze the impact of the number of measurements, the initial measurement set and the load fluctuation. Table \ref{tab:Status of Different Scenarios} illustrates the status of different scenarios. The difference between scenario 2 and scenario 1 is that in scenario 2, the initial measurement set is nodes $\{7,12,31,40,52,72,84,97,105,115\}$.
The settings of scenarios 3 and 4 are the same as those of scenario 1 except for the numbers of measurements are \textit{K} = $7$ and \textit{K} = $15$, with the initial measurement sets $\{22,23,24,26,27,28,29\}$, and $\{19,20,21,22,23,$ $24,26,27,28,29,30,31,32,33,34\}$ on each phase, respectively. The difference between scenario 5 and scenario 1 is that scenario 5 exhibits more significant load fluctuations. The corresponding factor that multiplies with the given load of scenario 5 is illustrated in Fig. \ref{fig:load_scenario5}. It can be observed that with the same number of the measurement, scenario 2 achieves a \textit{good system real-time assessment} (status 1 or 2) more quickly, with 3 time steps. This is because its initial measurement set is more evenly distributed, allowing it to gather more information about the system and build a better optimization problem (P1). Comparing scenario 1 with scenarios 3 and 4, which have 10, 7 and 15 measurements on each phase respectively, we observe that an increase in the number of measurements allows for faster attainment of a \textit{good system real-time assessment} (status 1 or 2). In scenario 5, we observe that although we achieve status 1 and obtain a \textit{good system real-time assessment} at time 9, at time 11 to 14 and time 23 to 24, we need to update the measurement set and cannot get a \textit{good system real-time assessment}. This is because in scenario 5, the load fluctuations are more significant, and the new measurement set obtained from the current time step may not sufficiently improve the lowest achievable voltage value $\underline{u}^{\phi,\text{lowest}}$ at the next time step. Nevertheless, compared to time 0, we observe that during time steps 11 to 14 and time steps 23 to 24, fewer time steps are required to reach status 1 or 2. This is due to a better initial measurement set at these times, which facilitates a more effective problem (P1).
\begin{table*}[htbp]
  \centering
  \setlength\tabcolsep{3pt}
  \caption{Status of Different Scenarios}
    \begin{tabular}{lrrrrrrrrrrrrrrrrrrrrrrrrrrrrrr}
    \hline
    \hline
    Time  & 0     & 1     & 2     & 3     & 4     & 5     & 6     & 7     & 8     & 9     & 10    & 11    & 12    & 13    & 14    & 15    & 16    & 17    & 18    & 19    & 20    & 21    & 22    & 23    & 24    & 25    & 26    & 27    & 28    & 29 \\
    \hline
    Scenario 1 & 0     & 0     & 0     & 0     & 0     & 0     & 0     & 0     & 0     & 0     & 2     & 2     & 0     & 2     & 1     & 1     & 1     & 1     & 2     & 1     & 1     & 2     & 2     & 2     & 1     & 1     & 2     & 2     & 2     & 1 \\
    Scenario 2 & 0     & 0     & 1     & 1     & 1     & 1     & 1     & 2     & 1     & 1     & 1     & 2     & 2     & 2     & 2     & 1     & 1     & 1     & 1     & 1     & 1     & 1     & 1     & 1     & 1     & 1     & 1     & 1     & 1     & 1 \\
    Scenario 3 & 0     & 0     & 0     & 0     & 0     & 0     & 0     & 0     & 0     & 0     & 0     & 0     & 0     & 0     & 1     & 1     & 1     & 1     & 1     & 1     & 1     & 1     & 1     & 1     & 1     & 1     & 1     & 1     & 1     & 1 \\
    Scenario 4 & 0     & 0     & 1     & 1     & 1     & 1     & 1     & 2     & 1     & 1     & 1     & 2     & 2     & 2     & 2     & 1     & 1     & 1     & 1     & 1     & 1     & 1     & 1     & 1     & 1     & 1     & 1     & 1     & 1     & 1 \\
    Scenario 5 & 0     & 0     & 0     & 0     & 0     & 0     & 0     & 0     & 0     & 1     & 1     & 0     & 0     & 0     & 0     & 1     & 1     & 1     & 2     & 2     & 1     & 1     & 2     & 0     & 0     & 1     & 1     & 2     & 1     & 2 \\
    \hline
    \hline
    \end{tabular}%
  \label{tab:Status of Different Scenarios}%
\end{table*}%

\section{Conclusion}

In this paper, we propose a real-time assessment approach of distribution power security through adaptive smart meter measurements. Considering the bandwidth limitations in the smart meter communications infrastructure, we formulate an optimization problem to determine the extreme achievable voltage magnitude across all nodes for a setting with limited measurements. We identify criteria under which we can certify that the system is operating securely, or where we can identify that we have found the nodes with the highest and/or lowest voltage magnitudes in the system.  
Further, we propose a strategy to adjust the set of smart meters where we obtain measurements in real time based on solutions from our optimization problems. This strategy is designed to help narrow the range of extreme achievable voltages, and ideally identify the true range of voltages in the system. Numerical analyses were tested using the modified three-phase unbalanced IEEE 123-bus distribution systems. The results illustrate that the proposed strategy can find better bounds on the extreme achievable voltage magnitude and enhance the real-time assessment, even when load fluctuations are present. The algorithm also successfully identifies the node with the most extreme voltage and obtain voltage magnitudes at this node. 
Moreover, test results illustrate that with more measurements and a better initial measurement set, it takes fewer time steps to achieve a good real-time assessment. If the load fluctuation is significant, it needs to update the measurement set more frequently to improve the real-time assessment. In future work, we want to investigate how to determine an optimal number of measurements in real time and how to deal with large load fluctuations.





{
\bibliographystyle{IEEEtran}
\bibliography{BIB}
}
\end{document}